\documentclass[11pt]{article}
\usepackage{amsmath,amssymb,color,epsfig,cite}
\usepackage{xcolor}
\usepackage{mathrsfs}

\usepackage{mathtools}

\usepackage{amsfonts}

\newcommand{\hoch}[1]{$\, ^{#1}$}

\makeatletter
\@addtoreset{equation}{section}
\makeatother

\newcommand{\be}{\begin{equation}}
\newcommand{\ee}{\end{equation}}
\newcommand{\bea} {\begin{eqnarray}}
\newcommand{\eea}{\end{eqnarray}}
\newcommand{\nn}{\nonumber}

\def\ft#1#2{{\textstyle{\frac{\scriptstyle #1}{\scriptstyle #2} } }}
\def\fft#1#2{{\frac{#1}{#2}}}

\def\0{{\sst{(0)}}}
\def\1{{\sst{(1)}}}
\def\2{{\sst{(2)}}}
\def\3{{\sst{(3)}}}
\def\4{{\sst{(4)}}}
\def\5{{\sst{(5)}}}
\def\6{{\sst{(6)}}}
\def\7{{\sst{(7)}}}
\def\8{{\sst{(8)}}}
\def\sst#1{{\scriptscriptstyle #1}}

\def\del{{\partial}}

\def\equal{{\,\,\,\stackrel{\raisebox{-1pt}{$\sst{*}$}}{=}\,\,\, }}

\def\cA{{{\cal A}}}

\def\im{{{\rm i\,}}}
\def\R{{\mathbb R}}

\def\bx{{{\mathrm x}}}

\def\bx {{\bf x}}

\def\ie{{ i.e.~}}

\def\ben{\begin{equation}}
\def\bea{\begin{eqnarray}}
\def\een{\end{equation}}
\def\eea{\end{eqnarray}}

\def \p {\partial}

\def\ft#1#2{{\textstyle{\frac{\scriptstyle #1}{\scriptstyle #2} } }}
\def\fft#1#2{{\frac{#1}{#2}}}

\begin{document}

\begin{flushright}
\hfill{UPR-1314-T\ \ \ \ MI-HET-761} 

\end{flushright}

\begin{center}
{\large {\bf Supergravity Black Holes, Love Numbers and Harmonic Coordinates}}

\vspace{15pt}
{\large M. Cveti\v c$^1$, G.W. Gibbons$^2$, C.N. Pope$^{3,2}$ and 
              B.F. Whiting$^{4}$ }

\vspace{15pt}

{\hoch{1}\it Department of Physics and Astronomy,\\
University of Pennsylvania, Philadelphia, PA 19104, USA}

\vspace{10pt}

\hoch{2}{\it DAMTP, Centre for Mathematical Sciences,
 Cambridge University,\\  Wilberforce Road, Cambridge CB3 OWA, UK}

\vspace{10pt}

\hoch{3}{\it George P. \& Cynthia Woods Mitchell  Institute
for Fundamental Physics and Astronomy,\\
Texas A\&M University, College Station, TX 77843, USA}

\hoch{4}{\it Department of Physics, P.O. Box 118440, University of Florida,
  Gainesville, FL 32611-8440}



\end{center}

\begin{abstract}

To perform realistic tests of theories of gravity, we need to be able to 
look beyond general relativity
and evaluate the consistency of alternative theories with observational 
data from, especially, gravitational wave detections using, for example, 
an agnostic Bayesian approach.  In this paper we further examine properties 
of one class of such viable, alternative theories, based on metrics 
arising from ungauged supergravity.  In particular, 
we examine the massless, neutral, minimally coupled scalar wave equation
in a general stationary, axisymmetric background metric such as that of 
a charged rotating black hole, when the scalar field is either time 
independent or in the low-frequency, near-zone limit, with a view to
calculating the Love numbers of tidal perturbations, and of
obtaining harmonic coordinates for the
background metric. For a four-parameter family of charged
asymptotically flat rotating
black hole solutions of ungauged supergravity theory known as
STU black holes, which includes Kaluza-Klein black holes
and the Kerr-Sen black hole as special cases, we find
that all time-independent solutions, and hence the harmonic
coordinates of the metrics, are identical to those of
the Kerr solution. In the low-frequency limit we find the scalar fields
exhibit the same $SL(2,R)$ symmetry as holds in the case of the Kerr solution.
We point out extensions of our results to a wider class of metrics,
which includes solutions of Einstein-Maxwell-Dilaton theory. 

\end{abstract}

\pagebreak

\tableofcontents
\addtocontents{toc}{\protect\setcounter{tocdepth}{2}}

\section{Introduction}

For now more than a century, tests of general relativity have typically 
attempted to show that predictions of the theory have been borne out by 
observation.  Thus, the default assumption in interpreting such
observations as gravitational waves by LIGO \cite{LIGOScientific:2016aoc} 
or the proposed space-borne LISA mission \cite{Maselli:2021men},
the apparent black hole shadow in M87 by EHT 
\cite{EventHorizonTelescope:2021dqv,Lima:2021las}, the black hole
at the centre of the Milky way \cite{Narang:2020bgo}, and X-ray emission
from black hole accretion discs \cite{Tripathi:2021rwb}, is in terms of
the Kerr metric. In the era to come there will be growing interest in 
attempting to show, or to exclude, the possibility that the results of 
observations may be better 
described by some alternative theory of gravity. Thus, in accessing the 
reliability of present-day interpretations,
or in looking for physics beyond the standard model,
it is important to compare with the predictions for 
alternative metrics; or, to adapt the terminology  
introduced in \cite{Damour:2007ap}, with \emph{Kerr metric foils}, that is,
asymptotically flat, axisymmetric, stationary metrics that are 
regular outside a non-singular event horizon\footnote{The ``foils'' of 
\cite{Damour:2007ap} were wormholes; ours are black holes.}.
To be useful for such purposes, one might require at the least
that:
\begin{itemize}
\item The Hamilton-Jacobi equation for null geodesics is Liouville 
integrable.   
\item The energy momentum tensor of all fields other than the metric satisfies 
acceptable 
positive-energy conditions. 
\item The metric is  a solution of a well-defined set of field equations,
 having a well-posed initial value problem. 
\item The propagation of time-dependent solutions is causal.  
\item The spacetime of the foil has positive energy.  
\end{itemize}

Although not essential, it is also highly desirable that the 
equations of motion, which might reasonably be assumed to contain only
massless scalar and vector fields in addition to the metric, 
may be derived from an action principle
and thus admit a Hamiltonian formulation and therefore a well-defined notion
of total energy and angular momentum. Moreover, 
progress  would be facilitated if, at least to some degree,
separability of the equations for linear perturbations of the solutions 
holds, analogous to the case for the Kerr metric, and also
that the analysis of magnetic fields around the foil solutions
be tractable. A final desirable requirement is that the foil
solutions have some degree of uniqueness, to provide
reassurance that predictions made using them are robust. 

With this in mind, we have previously studied the 
electrodynamics
\cite{Cvetic:2013roa}, the initial value problem \cite{Cvetic:2014vsa}, the
photon sphere and sonic horizons  \cite{Cvetic:2016bxi}, the equatorial
timelike geodesics \cite{Cvetic:2017zde}, and the
stability of massless, minimally-coupled scalar fields \cite{Cvetic:2019ekr}, 
in a remarkable
six-parameter family of exact rotating black hole
solutions  of ungauged supergravity theory known as \emph{STU
black holes} \cite{cy96b}.  These solutions admit a separation of variables
for the scalar wave equation \cite{cl97a,cl97b}.
In particular, this class of black hole spacetimes contains
as special cases: (i) the Kerr-Newman metric, (ii) its analogue in Kaluza-Klein
theory \cite{Rasheed:1995zv}, and (iii) the Kerr-Sen analogue in String Theory 
\cite{Sen:1992ua}.
The latter has frequently been used as a black hole foil in the 
literature \cite{EventHorizonTelescope:2021dqv, Lima:2021las, Narang:2020bgo}.

In this work, our main focus will again be on scalar fields in the STU 
supergravity
black hole backgrounds, since they
are easier to work with than the full gravitational perturbations while
still carrying the expectation that they will be indicative of behaviours
that would be manifested also in a more extensive analysis.  
One aim in the present paper is to extend our previous work to an 
examination
of tidal effects induced on rotating STU black holes by orbiting black holes
or neutron stars. These effects are characterized
by dimensionless numbers analogous to those introduced by
Love in his purely Newtonian study of tidal distortions of the
shape of the earth by the moon \cite{Love}. The definition
of tidal Love numbers\footnote{There has also been introduced a notion of 
dissipative Love
 numbers, about which we will say more below.} for  static $SO(d)$ invariant
asymptotically-flat vacuum black holes in $d+1$ dimensions
is given in \cite{Kol:2011vg}, where the reader will find an extensive 
list of references to earlier work.  Using the equations for static 
gravitational perturbations
of such metrics given in \cite{Ishibashi:2003ap}, the remarkable result
was obtained that the Love numbers vanish if $d=3$ but
are non-vanishing  for $d>3$. 

Recently \cite{LeTiec:2020spy,LeTiec:2020bos,Chia:2020yla}, it was shown 
that the vanishing 
of the tidal Love numbers also holds for the Kerr solution.
This may be seen \cite{Chia:2020yla} by taking
an appropriate limit of the Teukolsky equation \cite{Teukolsky:1973ha}
that governs scalar, spinor, vector and tensor perturbations.  
This achievement has sparked off attempts to provide an underlying 
explanation
for this remarkable phenomenon
\cite{Charalambous:2021mea,Charalambous:2021kcz,Hui:2020xxx,Hui:2021vcv}.

The paper is organised as follows. In section 2, we show by considering
an arbitrary stationary metric written in the form of a time fibration over
a spatial 3-dimensional base metric, that static (time independent) 
solutions of the scalar wave equation are governed by an equation that
depends only on the spatial 3-metric. In particular, this means that
not only for the charged rotating STU supergravity black holes, which we 
discuss explicitly, but also more generally in much broader classes of
charged black holes, the static scalar wavefunctions obey the same equation
in the charged metrics as they do in the uncharged ones.  In section 3 we
show that this means the static Love numbers for the scalar field in
the charged supergravity black holes are the same as those in an uncharged 
Kerr black hole.  We also comment on the distinction between the static
Love numbers, which therefore vanish, and the dynamical response that arises 
as a consequence of tidal interactions.  In section 4, we show that if one
looks at the properties of the scalar wave equation for non-zero but
 low frequency fields, then just as occurs in the Kerr background, one 
finds an $SL(2,\R)$ symmetry in the near-zone regime. 

   In section 5, we study the time-independent solutions of the massless 
scalar wave equation in two classes of static black hole backgrounds,
showing how in each case the 3-dimensional spatial metric is conformal
to the metric on 3-dimensional hyperbolic space.  This provides a geometrical
insight into the $SL(2,\R)$ symmetries of the static wavefunctions 
in these cases.  One of the classes of static black holes we study here are
4-charge black holes in STU supergravity.  The other class comprises 
charged black holes in Einstein-Maxwell-Dilaton (EMD) gravity. We consider these
balck holes for arbitrary values of the constant characterising 
the coupling of the dilaton to the Maxwell field.  Only in certain 
special discrete cases do these EMD black holes coincide with black holes in
the STU supergravity class.  

   In section 6, we apply some of our previous results to the 
construction of harmonic coordinates for the charged black hole metrics.
These are of interest because the Einstein equations then become
a semi-linear symmetric hyperbolic system, which can be useful for studying
certain mathematical properties of the solutions. They can also be used in
order to define the energy and momentum in terms of the Landau-Lifshitz
energy-momentum pseudotensor. The paper ends with conclusions in section 7.

\section{Scalar Wave Equation in Charged Rotating Black Hole Backgrounds}

\subsection{Charged from uncharged black holes}

  It was observed in \cite{chcvlupo4} that the charged rotating black 
holes of four-dimensional ungauged STU supergravity can be conveniently
constructed by starting from the neutral Kerr black hole as a seed 
solution,\footnote{To be more precise, the seed solution should be
taken to be the Kerr-NUT metric in general, with the seed 
NUT parameter eventually 
enabling the cancellation of a further NUT charge  
that can arise when certain combinations of sufficiently many charges
are introduced using the generating procedure.}
written in the time-fibred Kaluza-Klein form
\be
ds_4^2 = g_{\mu\nu}\, dx^\mu dx^\nu= 
 - e^{2U}\, (dt+\cA_i \,dx^i)^2 + e^{-2U}\, \gamma_{ij}\, dx^i dx^j\,,
\label{KKmet}
\ee
where the base metric $\gamma_{ij}$, the Kaluza-Klein vector $\cA_i$ and
the scalar $U$ depend only on the three 
spatial coordinates $x^i$.  By performing a 
Kaluza-Klein reduction of the STU supergravity theory 
on the time coordinate, using a general ansatz of the form (\ref{KKmet})
for the metric and 
\be 
\hat A_\mu dx^\mu  = A_i dx^i  + \chi\, dt \,,
\ee
for each of the four STU supergravity gauge potentials, the resulting
three-dimensional theory has an $O(4,4)$ global symmetry that can be used 
to introduce  
up to eight charges (independent 
electric and
magnetic charges for the four gauge fields), after lifting back to
four dimensions again.   

  An important point about this procedure of starting with a neutral
seed solution and introducing charges by using the global symmetries of
the three-dimensional reduced theory, is that the charge parameters 
enter the final four-dimensional metric only via the scalar function
$U$ and the Kaluza-Klein vector $\cA_i$ in eqn (\ref{KKmet}).  
This is because the metric $\gamma_{ij}$ in the 
three-dimensional reduced theory is invariant under the $O(4,4)$ global
symmetry.   In
other words, in the final four-dimensional charged solution, 
taking the form (\ref{KKmet}), 
the three-dimensional base metric $\gamma_{ij}$ is unchanged from the
form it took in the original seed solution.  

It is worth emphasizing that this procedure for generating charged solutions
from an uncharged seed solution can be applied much more generally than in
the specific case of the charged STU supergravity black holes we
are considering in this paper.  For any theory whose Kaluza-Klein reduction to 
three dimensions yields a theory with global symmetry group $G$, one
can (i) reduce from a stationary solution using the metric ansatz (\ref{KKmet}); 
(ii) act with the symmetry $G$; and then (iii) lift it back to 
four dimensions again.
Quite generally, since the three-dimensional metric is invariant under
$G$, the charge parameters in the lifted solution enter only via the
scalar function $U$ and the Kaluza-Klein vector $\cA_i$. Extensive discussions
of the three-dimensional global symmetries for rather general 
higher-dimensional starting points can be found, for example, in 
\cite{bregibmai} and \cite{cjlp3}.

As can easily be seen from eqn 
(\ref{KKmet}), the determinant of the four-dimensional metric 
$g_{\mu\nu}$ is related to that of the three-dimensional metric 
$\gamma_{ij}$ by $\sqrt{-g}= e^{-2U}\, \sqrt{\gamma}$, and so
\bea
\sqrt{-g} \,\Big(\fft{\del}{\del s_4}\Big)^2 = 
\sqrt{-g}\, g^{\mu\nu}\, \del_\mu\otimes \del_\nu=
-\sqrt{\gamma}\, e^{-4U}\, \del_t^2 + \sqrt{\gamma}\, \gamma^{ij}\, 
(\del_i - \cA_i\, \del_t)(\del_j -\cA_j\, \del_t)\,.
\eea
This means that the four-dimensional D'Alembertian wave operator 
$\square$ on scalar functions is given by
\bea
\square\Psi \equiv \fft1{\sqrt{-g}}\,\del_\mu\Big(\sqrt{-g}\, 
g^{\mu\nu}\, \del_\mu\Psi \Big)= - e^{-2U}\, \del_t^2\,\Psi 
  + e^{2U}\, 
\gamma^{ij}\, (\nabla_i -\cA_i\, \del_t)(\nabla_j -\cA_j\, \del_t)\Psi\,,
\label{KGgen}
\eea
where $\nabla_i$ is the covariant derivative in the three-dimensional
base metric $\gamma_{ij}$.  In particular, this
means that if the wavefunction\footnote{Note that in this paper we are 
using the
term ``wavefunction'' in a purely classical sense.} 
$\Psi(t,{\bx})$ is taken to be 
independent of time, $\Psi(t,{\bx})= \Psi({\bx})$, then
\bea
\square\Psi({\bx}) = e^{2U}\, \gamma^{ij}\, \nabla_i\nabla_j \Psi({\bx})
= \fft{e^{2U}}{\sqrt{\gamma}}\, 
\del_i\Big(\sqrt{\gamma}\, \gamma^{ij}\, \del_j\, 
\Psi({\bx})\Big)\,.\label{KGstatic}
\eea
Thus a time-independent solution of the massless wave equation
obeys
\bea
\gamma^{ij}\, \nabla_i\nabla_j \Psi({\bx})=0\,,
\eea
and this depends only on the metric $\gamma_{ij}$ of the 
three-dimensional base. As already observed, this is identical in the 
original neutral
seed solution and in the charged solution.

\subsection{Charged rotating black holes in STU supergravity}

  Written in the Kaluza-Klein form (\ref{KKmet}), the neutral Kerr
solution is given by the three-dimensional base metric 
\bea
\gamma_{ij}\, dx^i dx^j &=& 
(\rho^2 - 2M r)\Big(\fft{dr^2}{\Delta} + d\theta^2\Big) + 
      \Delta\,\sin^2\theta\, d\phi^2\,,\\
\rho^2&=&r^2+a^2\, \cos^2\theta\,,\qquad \Delta = r^2+a^2-2M r\,,
\label{3base}
\eea
together with the vector $\cA_i$ and scalar $U$: 
\bea
\cA_i\, dx^i = \fft{2M a \, r\,\sin^2\theta}{\rho^2 -2M r}\, d\phi\,,\qquad 
e^{2U}= 1 -\fft{2M r}{\rho^2}\,.\label{seedcAvarphi}
\eea
(Note that we can write $\cA_i dx^i= a (e^{-2U} -1)\, 
\sin^2\theta d\phi$ in the
seed solution.)

   Charged rotating STU black holes carrying just four charges in total
were obtained originally in \cite{cy96b}, and were obtained in
the Kaluza-Klein formulation in 
\cite{chcvlupo4}; for these solutions, 
the metric is given by (\ref{KKmet}) with the base
3-metric again given by (\ref{3base}), and the 1-form $\cA_\1$ and
scalar $U$ now given by
\bea
\cA_i\, dx^i &=& 
\fft{2M a \, [r\, \Pi_c -(r-2M)\, \Pi_s]\sin^2\theta}{\rho^2 -2M r}\, d\phi
\,,\qquad
e^{2U}= \fft{\rho^2}{W}\,\Big(1 -\fft{2M r}{\rho^2}\Big) \,,
\label{cAvarphi}
\eea
where
\bea 
W^2 &=& r_1 r_2 r_3 r_4 +
2a^2\,\Big[ r^2 + Mr \sum_i s_i^2 + 4M^2 \Pi_s(\Pi_c-\Pi_s) -
  2M^2\sum_{i<j<k} s_i^2 s_j^2 s_k^2\Big]\,\cos^2\theta\nn\\
&& + a^4\, \cos^4\theta\,,\label{Wdef}\\
r_i&=& r + 2M s_i^2\,,\quad s_i=\sinh\delta_i\,,\quad c_i=\cosh\delta_i
\,,\quad \Pi_s= s_1 s_2 s_3 s_4\,,\quad \Pi_c=c_1 c_2 c_3 c_4\,.\nn
\eea
As already remarked, the charge parameters (the boost parameters $\delta_i$)
enter only in the expressions for $\cA_i$ and $U$, and thus the
scalar wave operator for time-independent wavefunctions, given 
by (\ref{KGstatic}), remains
unchanged from that for the original Kerr metric.\footnote{Note that for the 
particular combination of four charges that are introduced in this
example, it suffices to take just the Kerr metric as the seed solution,
since no NUT charge is generated in this case.}

  More generally, the full eight-charge rotating black hole family
of solutions was obtained using these methods in 
\cite{chowcomp8}.\footnote{Because of the existence of an 
$SL(2,\R)^3$ global symmetry of the STU theory in four dimensions,
it actually suffices from the point of view of generality
to construct a five-charge solution, since the
remaining three charges can then be introduced in four dimensions
by acting with the $U(1)^3$ maximal compact subgroup of
$SL(2,\R)^3$.  This technique was employed for the case of the
static STU black holes in \cite{cvetyoum5static}.}  As
can be seen from the expressions given there, the charge parameters
again enter only in the Kaluza-Klein vector $\cA_i$ and the scalar $U$ in
the metric (\ref{KKmet}), and so again the massless wave equation for
the case of time-independent wavefunctions is independent of the
charge parameters.

\section{Time-independent Scalar Wavefunctions, and Love Numbers}

  Consider a time-independent solution $\Psi(r,\theta,\phi)$ of the massless 
scalar wave
equation in the background of a charged rotating STU supergravity
black hole.  From eqn (\ref{KGstatic}) and (\ref{3base}),  it will obey
\bea
\del_r(\Delta\del_r\Psi) + \fft1{\sin\theta}\,\del_\theta(\sin\theta\,\Psi)
   - \fft{(\rho^2 - 2Mr)\, m^2}{\Delta\,\sin^2\theta}\, \Psi=0\,.
\eea
From the expressions for $\rho$ and $\Delta$ in eqns (\ref{3base}) 
we have
\bea
\fft{\rho^2 - 2Mr}{\Delta\,\sin^2\theta} = 
\fft1{\sin^2\theta} -\fft{a^2}{\Delta}\,,
\eea
and so for factorised solutions with $\Psi(r,\theta,\phi)=
R(r)\,S(\theta)\, e^{\im m\phi}$, the massless 
scalar wave equation separates, giving
\bea
\fft{1}{\sin\theta}\, \del_\theta (\sin\theta\, \del_\theta S) +
  \Big[ \lambda -\fft{m^2}{\sin^2\theta}\Big]\, S=0\,,\label{Seqn}
\eea
implying that $S(\theta)$ is just
the associated Legendre function $P_\ell^m(\cos\theta)$, and the
separation constant $\lambda$ is
\bea
\lambda=\ell(\ell+1)\,,\qquad \ell=0,1,2,\cdots\,,
\eea
and therefore the radial equation is
\bea
\del_r(\Delta\,\del_r R) + \fft{a^2\,m^2}{\Delta}\, R-\ell(\ell+1)\, R=0\,.
\label{Reqn}
\eea

\subsection{Love numbers}

    In non-relativistic gravity the Newtonian potential $U$ of a tidally 
distorted body of mass $M$ and mean radius $R$ is given in 
spherical coordinates by 
\cite{PW,LeTiec:2020spy}
\ben\label{Love-eq}
U= \frac{M}{r} - \sum_{\ell,m} \frac{(\ell-2)!}{\ell!} E_{\ell m} r^\ell
\left[ 1 + 2  k_\ell \left( \frac{R}{r}\right)^{2\ell+1}   
\right]Y_{\ell m} (\theta ,\phi) \,, 
\een
where $Y_{\ell m}$ are spherical harmonics,
$E_{\ell m}$ are  a measure of the moments of an external
tidal field 
and $k_\ell E_{\ell m}  R^{2\ell+1}$ is a measure of the deformation  of
the gravitational field of the body. The quantity $R$ is included on 
dimensional grounds,
and renders the coefficients $k_\ell$, known as Tidal Love Numbers, or TLNs,
dimensionless. The name Love refers to the elastician
Augustus Edward Hough Love who introduced the $k_\ell$ 
coefficients in \cite{Love}.
The Love numbers provide a charactarization of the elasticity or
rigidity of the body, with larger Love numbers corresponding to
greater elasticity.  However, as Love noted  \cite{Love}, 
tidal forces are often dynamical, and (\ref{Love-eq}) represents only 
the static part of an elastic body's response to tidal deformation.

   In general relativity the situation is obviously more complicated and, for black holes, even more so.  In particular, if the Love numbers are zero this 
does not indicate, as has sometimes been supposed, that a black hole has 
no response to an external tidal field.  In fact, early studies of 
perturbations of black holes found that their surrounding spacetime would 
respond with quasi-normal ringing \cite{Press:1971wr, Chandrasekhar:1975zza}, 
and York \cite{York:1983zb} showed that these actually resulted in changes 
in the area of the event horizon of a black hole.  In numerical relativity 
it was later demonstrated that during black hole mergers, cross-sections of 
the event horizon could become singular \cite{Matzner:1995ib}.  Thus, the 
static Love numbers being zero for black holes represents only part of 
the full story.

   In examining the Love numbers 
of the Kerr solution, the approach usually taken has been to focus 
on the Teukolsky equations \cite{Teukolsky:1973ha}, 
which govern gauge invariant parts of the Weyl tensor arising from 
gravitational perturbations of the metric, or some
equivalent formulation.
These equations are necessarily rather complicated, and in any case so far
their generalizations are not available for all of the
rotating black holes we wish to consider. However, on the basis
of our results, some insight may be gained
from the behaviour of static, i.e. time-independent, solutions
of the wave equation $\Box  f=0$ where $\Box$ is the
covariant D'Alembertian
of any metric of the form (\ref{KKmet}).

 In \cite{Charalambous:2021kcz}, static Love numbers, which
   determine the  response to time-independent external fields, were
found to vanish in four-dimensional Einstein theory both
for spherical and spinning black holes (see also
\cite{Binnington:2009bb,Fang:2005qq,Damour:2009vw,Kol:2011vg,
Hui:2020xxx,Chia:2020yla,Charalambous:2021mea}).
Although this might appear to be at variance with
\cite{LeTiec:2020spy,LeTiec:2020bos}, 
it is clear from \cite{Chia:2020yla} that dynamics is at the root 
of this apparent discrepancy.  Specifically, while the static Love number 
is indeed zero for the Kerr black hole, there is a dissipative response 
(in the Weyl tensor) proportional to a superradiance factor, namely 
$$-\im (m\Omega_H-\omega)\left[\frac{2Mr_+}{r_+-r_-}\right]
 \nu_{\ell m}^{\rm Kerr}$$
which, except for the axisymmetric $m=0$ mode, does not vanish in the 
zero-frequency limit (nor, indeed, in Schwarzschild with non-zero frequency).  
Here $r_-$ and $r_+$ label the inner and outer horizons, 
$\Omega_H$ is the angular momentum of the outer horizon, and the 
$\nu_{\ell m}$ are new, {\it dissipative}, Love numbers.  
Note that the sign of 
this term changes in the superradiant regime.  
This discrepancy is specifically addressed in \cite{LeTiec:2020bos}, and
in its discussion of
the papers \cite{Goldberger:2020fot} and \cite{Charalambous:2021mea}.

As implied by our earlier discussion, 
  the radial equation arising from the separation of variables for
time-independent solutions of the massless scalar wave equation in 
the background of any charged rotating STU supergravity black hole
is given by (\ref{Reqn}).  Since this is identical in form
to the corresponding radial equation in the Kerr black hole background,
the same analysis given in \cite{Hui:2021vcv,Charalambous:2021kcz} 
carries over identically to the charged
supergravity black hole cases.

\section{Near-Zone $SL(2,\R)$ Symmetry}

  It was shown in \cite{klemm} that the massless scalar wave equation for
low-frequency wavefunctions 
in the Schwarzschild geometry exhibits a
``hidden'' $SL(2,\R)$ symmetry.  This observation was subsequently 
extended to more general black hole backgrounds, including the
Kerr black hole in \cite{Charalambous:2021kcz}.  In this section, we show that
the hidden $SL(2,\R)$ symmetry is present also for the low-frequency
massless wave equation in the background 
of the 4-charge rotating STU supergravity black holes.  This is 
noteworthy because, unlike the zero-frequency results that we discussed 
earlier, which were insensitive to any of the details of the charge
parameters, here the low-frequency massless wave equation does
involve dependence on the charge parameters.

  Considering now time-dependent massless scalar wavefunctions 
$\Phi=R(r)\, S(\theta)\,
e^{-\im\omega t+ \im m\phi}$, one can see from the expressions in
 eqns (\ref{3base}), (\ref{cAvarphi}) and (\ref{Wdef}) that the 
scalar wave operator (\ref{KGgen}) separates, giving the angular
equation 
\bea
\fft1{\sin\theta}\,\del_\theta(\sin\theta\,\del_\theta S) +
\Big[\lambda+ 
  a^2\, \omega^2\, \cos^2\theta -\fft{m^2}{\sin^2\theta}\Big]\, S=0\,,
\eea
and the radial equation 
\bea
\del_r(\Delta\del_r R) + (V_0 + V_1) R=\lambda R\,,\label{eqrad}
\eea
where
\bea
V_0&=& \fft{4M^2\,(\Pi_c\, r_+ + \Pi_s\, r_-)^2}{\Delta}\, 
\Big[(\omega - m\,\Omega)^2 - 4m \omega \Omega
\,\fft{r-r_+}{r_+ - r_-}\Big]\,,\label{V0}\\
V_1 &=& \fft{2 a m\omega\, M\, (\Pi_c +\Pi_s)}{
                               \kappa\,(r-r_-)\,(\Pi_c\, r_+ + \Pi_s\, r_-)} +
  \fft{8\omega^2\, M^3(\Pi_c^2-\Pi_s^2)}{r-r_-} +
  \omega^2(b_0 + b_1\, r + r^2)\,,\label{V1}\\
b_0&=& 4M^2 \Big(1+\sum_i s_i^2 + \sum_{i<j} s_i^2 s_j^2\Big)\,,\qquad
b_1 = 2M \Big(1+\sum_i s_i^2\Big)\,.
\eea
Here $\kappa$, the surface gravity of the outer horizon, 
and $\Omega$, the angular velocity
of the outer horizon, are given by
\bea
\kappa=\fft{r_+ - r_-}{4M (\Pi_c\, r_+ + \Pi_s\, r_-)}\,,\qquad
\Omega= \fft{a}{2M (\Pi_c\, r_+ + \Pi_s\, r_-)}\,.
\eea

The decomposition of the potential in the radial equation as the sum
of the two terms $V_0$ and $V_1$ is motivated by the analysis in
\cite{Charalambous:2021kcz} for the case of the Kerr metric.  The term $V_0$
contains all the terms that contribute at leading order in the 
near-zone region defined by
\bea
\omega\, (r-r_+) <<1\,,\label{nearzone}
\eea
while the term $V_1$ is of subleading order in a near-zone 
expansion.\footnote{As noted in \cite{Charalambous:2021kcz}, 
there is some arbitrariness
in the choice of the decomposition into the terms $V_0$ and $V_1$.  The
important point is that the terms in $V_0$ with $(r-r_+)$ in the
denominator are the dominant ones in the near zone. The last term in 
the expression for $V_0$ in (\ref{V0}) (for which the $(r-r_+)$ factor 
coming from $\Delta=(r-r_+)(r-r_-)$ in the denominator, is cancelled
by the factor in the numerator), could just as well be assigned to
$V_1$.  It is included in $V_0$ in order to give a precise
formulation of the near-zone equation that exhibits the $SL(2,\R)$ 
symmetry, in eqn (\ref{eqradnear}) as seen below.}

  Following \cite{Charalambous:2021kcz}, 
and making the appropriate changes for our
case, we now define $SL(2,\R)$ generators as follows:
\bea
L_0 &=& \kappa^{-1}\, \del_t\,,\nn\\
L_{\pm1} &=& e^{\pm \kappa t}\, \Big(
  \mp \Delta^{1/2}\, \del_r + \kappa^{-1}\, \del_r(\Delta^{1/2})\, \del_t
  + \fft{a}{\Delta^{1/2}}\, \del_\phi\Big)\,.\label{sl2rgen}
\eea
These obey the $SL(2,\R)$ algebra 
\bea
[L_m,L_n] =(m-n)\, L_{m+n}\,,\qquad -1\le m\le 1\,,\quad -1\le n\le 1\,. 
\label{sl2ralg}
\eea
Defining the quadratic $SL(2,\R)$ Casimir operator ${\cal C}_2 \equiv L_0^2 -
\ft12(L_{-1}\, L_1 + L_1\, L_{-1})$, the scalar wave equation in the
near zone can (\!\ie with the $V_1$ term suppressed) be written as
\bea
{\cal C}_2\, \Phi= \lambda\,\Phi\,,\label{KGnear}
\eea
with the radial function $R$ satisfying the near-zone equation
\bea
\del_r(\Delta\del_r R) + V_0\, R=\lambda R\,,\label{eqradnear}
\eea
Thus, the near-zone massless scalar wave equation exhibits an $SL(2,\R)$
symmetry in the 4-charge rotating black hole background, extending the
previous findings for the Schwarzschild \cite{klemm} and Kerr
\cite{Charalambous:2021kcz} black holes.

\section{Time-independent Scalar Fields in Static Black Hole Backgrounds}

  It was observed in \cite{Hui:2020xxx} that in the case of the
Schwarzschild black hole background, the symmetries of the
time-independent scalar wavefunctions, reflected in the tower of
ladder operators that related the the solutions with different
values of $\ell$ in the decomposition in spherical harmonics 
$Y_{\ell m}(\theta,\varphi)$, were related to the fact that the
massless time-independent wave operator was conformally related to
that on three-dimensional hyperbolic space.  In this section, we demonstrate
that this feature extends to two broad classes of static black hole
solutions, namely those in a general Einstein-Maxwell-Dilaton
(EMD) theory, and to the static 4-charge black holes of STU supergravity.

\subsection{Static Einstein-Maxwell-Dilaton black holes}

  These black holes are solutions of the theory described by the
Lagrangian
\bea
{\cal L} = \sqrt{-g}\, \Big[R - 2(\del\phi)^2 - e^{-2a\phi}\, F^2\Big]\,,
\eea
where the dimensionless constant, $a$, can be arbitrary.
The static black hole solutions were obtained in \cite{gibmae}.  In a
convenient parameterisation described in \cite{Cvetic:2016bxi}, the metrics
are given in terms of an isotropic radial coordinate $\rho$ by
\bea
ds^2 = -e^{2U}\, dt^2 + \Phi^4\, (d\rho^2 + \rho^2\, d\Omega^2)\,,
\eea
where
\bea
e^{2U} &=& V^2\, W^{\fft{2(1-a^2)}{1+a^2}}\,(CD)^{-\fft{2}{1+a^2}}\,,
\qquad
\Phi^2 = (CD)^{\fft1{1+a^2}}\, W^{\fft{2a^2}{1+a^2}}\,,\nn\\
C&=& 1+\fft{u^2}{\rho}\,,\qquad 
D= 1+\fft{v^2}{\rho}\,,\qquad
 W=1+ \fft{uv}{\rho}\,,\qquad
V=1 -\fft{u v}{\rho}\,,
\eea
where $u$ and $v$ are constants parameterising the mass and the charge. 

  It can be seen that 
\bea
\sqrt{-g}\, g^{\rho\rho} = \rho^2\, VW\, \sin\theta\,,\qquad
\sqrt{-g}\, g^{\theta\theta}= VW\,\sin\theta\,,\qquad
\sqrt{-g}\, g^{\varphi\varphi}= \fft{VW}{\sin\theta}\,,
\eea
and therefore a time-independent solution $\Psi$ of the massless scalar wave
equation obeys
\bea
\del_\rho(\rho^2\, VW \,\del_\rho\Psi) + VW\, \nabla_{(\theta,\varphi)}^2
\Psi=0\,,\label{EMDstatic}
\eea
where $\nabla_{(\theta,\varphi)}^2=
\csc\theta\,\del_\theta(\sin\theta\,\del_\theta) + 
\csc^2\theta\, \del_\varphi^2$ is the scalar Laplacian on the unit sphere.
In terms of the standard radial coordinate $r$ defined by
\be
r= \rho C D= \rho + u^2+v^2 + \fft{u^2 v^2}{\rho}\,,
\ee
the outer and inner horizons are located at $r_\pm = (u\pm v)^2$, and
defining $\bar\Delta=(r-r_+)(r-r_-)$ we have
\bea
\bar\Delta= \rho^2\, V^2\, W^2\,.
\eea
The time-independent wave equation (\ref{EMDstatic}) becomes
\bea
\del_r(\bar\Delta\, \del_r\Psi) + \nabla_{(\theta,\varphi)}^2
\Psi=0\,,
\eea
and this can be seen to be the Laplace equation in the 3-metric
\bea
ds_3^2 = dr^2 + \bar\Delta\, d\Omega^2\,.\label{3metunt}
\eea
Following the same steps as were described in \cite{Hui:2020xxx} in the
case of the Schwarzschild metric, we find here for the static EMD
black holes that the conformally-rescaled metric $d\tilde s_3^2=\Omega^2\,
ds_3^2$, with $\Omega=L^2/\bar\Delta$, is given in terms of the isotropic
radial coordinate $\rho$ by
\be
d\tilde s_3^2 = \fft{L^4}{(\rho^2-u^2 \,v^2)^2}\, 
(d\rho^2 + \rho^2\, d\Omega^2)\,.
\ee
This can be recognised as the homogeneous metric on three-dimensional 
hyperbolic space.

\subsection{Static 4-charge STU supergravity black holes}

  These can be conveniently written, in terms of an isotropic radial
coordinate, in the form \cite{Cvetic:2016bxi}
\bea
ds^2 &=& -\Pi^{-1/2}\, f_+^2\, f_-^2\, dt^2  + \Pi^{1/2}\, 
(d\rho^2 + \rho^2\, d\Omega^2)\,,\nn\\
\Pi &=& \prod_{I=1}^4 C_I\, D_I\,,\quad
f_\pm = 1 \pm \fft{m}{2\rho}\,,\quad
C_I= 1 + \fft{m e^{2\delta_I}}{2\rho}\,,\quad
D_I = 1 + \fft{m e^{-2\delta_I}}{2\rho}\,,
\eea
where the constants $m$ and $\delta_I$ parameterise the mass and the four
charges.  The massless wave equation for a time-independent function
$\Psi$ is therefore given by
\bea
\del_\rho(\rho^2\, f_+\, f_-\, \del_\rho\Psi)+ f_+\, f_-
\nabla_{(\theta,\varphi)}^2\Psi=0\,.  \label{4chstat}
\eea

   In terms of the radial coordinate $r$ defined by
\be
r= \rho + m +\fft{m^2}{4\rho}\,,
\ee
and defining $\bar\Delta\equiv r(r-2m)=\rho^2\, f_+^2\, f_-^2$, eqn 
(\ref{4chstat}) becomes
\bea
\del_r (\bar\Delta\, \del_r\Psi) + \nabla_{(\theta,\varphi)}^2\Psi=0\,.
\eea
This is the Laplace equation in the metric $ds_3^2= dr^2 +\bar\Delta\,
d\Omega^2$.  The conformally-rescaled metric $d\tilde s_3^2=
\Omega^2\, ds_3^2$, with $\Omega=L^2/\bar\Delta$, is recognisable as
the metric
\bea
d\tilde s_3^2 = \fft{L^4}{(\rho^2-\fft{m^2}{4})^2}\, 
(d\rho^2 + \rho^2\, d\Omega^2) \,,
\eea
on three-dimensional hyperbolic space, 
when written in terms of the isotropic coordinate
$\rho$.

  In summary, we have seen that both in the case of the static EMD black
holes and the static 4-charge STU black holes, the massless time-independent
scalar wave equation reduces to the same form as was found in 
\cite{Hui:2020xxx} in the case of the Schwarzschild metric.  Thus the
construction of the time-independent wavefunctions in terms of 
ladder operators proceeds in the same way as was seen in 
\cite{Hui:2020xxx}, with the underlying conformally-related hyperbolic
metric providing a geometric understanding for the ladder structure.

   In all the static backgrounds considered, the time-independent solutions
of the massless scalar wave equation obey the Laplace equation
\be
g^{ij}\, \nabla_i\nabla_j\,\Psi\equiv  \nabla^2\,\Psi=0 \,,
\ee
in a 3-metric of the form (\ref{3metunt}) with $\bar\Delta$ given by
\be
\bar\Delta= (r-r_+)(r-r_-)\,.\label{Deltagen}
\ee
In the conformally-rescaled metric $d\tilde s_3^2= \Omega^2\, ds_3^2$
with $\Omega=L^2/\bar\Delta$, the conformally-invariant three-dimensional 
scalar operator $(\widetilde\nabla^2 -\ft18 \widetilde R)$ is related to
the corresponding operator in the untilded metric by
\bea
(\widetilde\nabla^2 -\ft18 \widetilde R)\,\widetilde\Psi=
  \Omega^{-5/2}\, (\nabla^2-\ft18 R)\,\Psi\,,\label{d3conf}
\eea
where
\bea
\widetilde\Psi = \Omega^{-1/2}\, \Psi\,.
\eea
As can be easily verified, when $\bar\Delta$ is given by (\ref{Deltagen}) 
the Ricci scalar in the untilded metric 
(\ref{3metunt}) is given by
\be
R=\fft{(r_+ - r_-)^2}{2\bar\Delta^2}\,,
\ee
while the Ricci scalar of the tilded metric 
$d\tilde s_3^2 = \Omega^2\, ds_3^2 =  (L^4/\bar\Delta^2)\, ds_3^2$ is given by
\bea
\widetilde R= -\fft{3(r_+ - r_-)^2}{2 L^4}\,.
\eea
Thus it follows that $-\ft18 \widetilde R + \ft18 \Omega^{-2}\, R=
-\ft16 \widetilde R$, and so from eqn (\ref{d3conf}) we see that if
$\Psi$ obeys $\nabla^2\Psi=0$ then $\widetilde\Psi$ obeys
\bea
(\widetilde\nabla^2 - \ft16 \widetilde R)\,\widetilde\Psi=0\,.\label{d4conf}
\eea
In other words, in all the static metrics we
are considering here the time-independent solutions of the
massless scalar wave equation in the four-dimensional static black hole
background are conformally related to solutions of
eqn (\ref{d4conf}) in three-dimensional hyperbolic space.  Note that,
as was observed in \cite{Hui:2020xxx} in the case of the Schwarzschild
metric, 
the coefficient of the Ricci scalar in this equation is the one associated
with the conformally-invariant scalar wave equation of four dimensions, 
not three.  As was discussed in \cite{Hui:2020xxx}, the fact that the
metric $ds_3^2$ whose Laplacian gives the time-independent scalar
wavefunctions is conformal to the $d\tilde s_3^2$ hyperbolic
metric, which has $SO(3,1)$ as its symmetry group, provides one
way to understand the 
$SL(2,R)$ symmetry and ladder structure of the wavefunctions. In the
conformal frame of the hyperbolic metric the conformally-rescaled 
wavefunctions $\widetilde\Psi$ are eigenfunctions of the Laplace
operator $\widetilde\nabla^2$ shifted by the constant $-\ft16\widetilde R$.

\section{Harmonic Coordinates}

 We turn, in this section, to an application of our previous results, enabling the
 construction of harmonic coordinates, also known as wave coordinates.
Let
\ben
\frak{g} ^{\mu \nu}  = \sqrt{-\det g _{\alpha \beta}}\,  g^{\mu \nu }
\label{Theta}   \,.
 \een
Then, by definition, a harmonic coordinate chart,
$ x^\mu = (x^0, x^1, x^2, x ^3) = (x^0, x^i )$,
is one  in which the De Donder gauge condition   
\ben
\p_\mu\, \mathfrak{g} ^{\mu \nu} =  0 \,,     \label{defn1}
\een
holds. It follows that in harmonic coordinates the D'Alembertian 
of any scalar function $f$ is given by 
\ben
 \Box f \equal  g^{\mu \nu} \p_\mu \p_\nu f \label{Box} \,
\een
where $\!\!\!\equal\!\!\!$ means ``equal in harmonic coordinates".  In particular, harmonic coordinates satisfy
\ben\label{boxeq}
\Box\, x^\alpha =0 \,.
\een
The main interest of harmonic coordinates is that in  
these coordinates the Einstein equations
become a  semi-linear symmetric hyperbolic system,
since the highest derivative term has the same
form as the right-hand side of (\ref{Box}), with $f$ representing
each component of the metric $g_{\alpha \beta }$ 
\cite{Choquet-Bruhat:2009xil}. They also permit a reformulation of
the Einstein equations and the definition of total energy and momentum
in terms of the Landau-Lifshitz energy momentum pseudo-tensor \cite{PW}.

   It is a standard result that for the Schwarzschild metric, 
a set of harmonic coordinates is provided by {\cite{Fromholz:2013hka,PW}
\ben
x^\alpha = \Big( t, (r-M) \sin \theta \cos \phi, 
  (r-M) \sin \theta \sin \phi, (r-M) \cos \theta\Big) \,.
\een
This result has been extended to the case of the Reissner-Nordstr\"om 
black hole in \cite{He:2014nna}, the Kerr black hole in \cite{Jiang:2014zz}, and the Kerr-Newman black hole in \cite{Lin:2014laa}.
In view of the universal form of the scalar D'Alembertian (\ref{KGgen}) 
for the STU charged black holes we are considering in this paper, it
is evident that the procedure found in \cite{Jiang:2014zz} for constructing
harmonic coordinates in the Kerr geometry can be immediately carried
over to the general case of the charged rotating STU black holes.  Thus
we define a new azimuthal coordinate $\tilde\phi$ by setting
\be
\tilde\phi= \phi +  \int^r \fft{a\, dr'}{\Delta(r')}\,,
\ee
and then it can be seen that the coordinates $x^\alpha$ defined by
\bea
x^\alpha =\Big(t, [(r-M) \cos\tilde\phi -a \sin\tilde\phi]
\, \sin\theta,[(r-M) \sin\tilde\phi + a \cos\tilde\phi]\, \sin\theta,
  (r-M)\, \cos\theta\Big)
\label{chargedharmonic}
\eea
are harmonic in the charged black hole metric (\ref{KKmet}), where 
the base 3-metric $\gamma_{ij}$ is given by that of the Kerr seed metric
in eqn (\ref{3base}).  
In any stationary, axisymmetric metric for which
all metric components are independent of both $\phi$ and $t$, and such that
$g^{t \theta  }$, $g^{t r}$, $g^{\phi  \theta}$,  and $g^{\phi r}$
all vanish, so that $g^{\phi t} = g^{t \phi}$ gives the only non-vanishing
cross term,
then $t$ (and, in fact, $\phi$ too) is a harmonic function, and we can seek the spatial harmonic coordinates by looking only at the time independent part of (\ref{boxeq}).
Thus, irrespective of ${\cal A}_i$ and $U$, the coordinates $x^\alpha$ given in eqn 
(\ref{chargedharmonic}) satisfy $\square\, x^\alpha=0$.  
It follows that,
in complete generality, eqn (\ref{chargedharmonic}) provides a set of
harmonic coordinates for all charged rotating black holes in the class
(\ref{KKmet}) with the Kerr seed 3-metric.

\section{Conclusions}

We have studied the charged, asymptotically-flat, rotating 
black hole solutions of ungauged STU supergravity, which include 
Kaluza-Klein black holes
and Kerr-Sen
black holes as special cases, and we find that the 
time-independent solutions of the
massless scalar wave equation are identical to those of the Kerr solution. 
This implies that the Love numbers for scalar perturbations, which
had previously been shown to vanish in the Kerr background, vanish also for
all these charged black holes.  The harmonic coordinates for the
charged black holes also coincide with those for the Kerr metric.
In the low-frequency
limit, we find the scalar fields exhibit the same $SL(2, R)$ symmetry as
holds in the case
of the Kerr solution. We have pointed out extensions of our results to
a wider class of black-hole metrics, including solutions of 
Einstein-Maxwell-Dilaton theory.

\section*{Acknowledgements}

MC is supported in part by DOE 
Grant Award de-sc0013528, the Fay R. and Eugene L. Langberg Endowed Chair, and
Slovenian Research Agency No. P1-0306.
CNP is supported in part by DOE grant DE-FG02-13ER42020.  
BFW is supported in part 
by NSF grant PHY 1607323,
and the Max Planck Institute for Gravitational Physics (Albert Einstein Institute), Potsdam, Germany.

\end{document}